\documentstyle[twocolumn]{article}
\def\be{\begin{equation}}
\def\ee{\end{equation}}
\def\bear{\be\begin{array}}
\def\eear{\end{array}\ee}
\def\bea{\begin{eqnarray}}
\def\eea{\end{eqnarray}}

\def\baselinestretch{1}
\begin{document}
\catcode`@=11
\newtoks\@stequation
\def\subequations{\refstepcounter{equation}%
\edef\@savedequation{\the\c@equation}%
  \@stequation=\expandafter{\theequation}
  \edef\@savedtheequation{\the\@stequation}
  \edef\oldtheequation{\theequation}%
  \setcounter{equation}{0}%
  \def\theequation{\oldtheequation\alph{equation}}}
\def\endsubequations{\setcounter{equation}{\@savedequation}%
  \@stequation=\expandafter{\@savedtheequation}%
  \edef\theequation{\the\@stequation}\global\@ignoretrue

\noindent}
\catcode`@=12
\begin{titlepage}
\title{{\bf "Soft" Phenomenology 
}\thanks{Based on a talk given at 
`International Europhysics Conference on High--Energy Physics',
Jerusalem (Israel), August 1997.}}
\author{{\bf C. Mu\~noz}
\thanks{Research supported in part by: the CICYT, under
contract AEN93-0673; the European Union,
under contracts CHRX-CT93-0132 and
SC1-CT92-0792; the KOSEF, under the Brainpool program.}\\
\hspace{3cm}\\
{\small Departamento de F\'{\i}sica 
Te\'orica C--XI}\\
{\small Universidad Aut\'onoma de Madrid,
Cantoblanco, 28049 Madrid, Spain} \\
{\small e--mail: cmunoz@bosque.sdi.uam.es}\\
\centerline{and} \\
{\small 
Department of Physics}\\
{\small Korea Advanced Institute of Science and Technology}\\ 
{\small Taejon 305--701, S. Korea}} 
\date{}
\maketitle
\def\baselinestretch{1.15}
\begin{abstract}
\noindent

A review of the soft supersymmetry--breaking parameters and the
$\mu$ term arising in superstring models is
performed paying special attention to their phenomenological 
implications. 
In particular, the violation of the scalar mass universality which
may lead to dangerous 
flavor changing neutral current
phenomena and the existence of charge and color breaking minima are discussed.
Finally, 
quadratically--divergent loop effects
by pure supergravity interactions on soft parameters and the $\mu$ term 
are also reviewed and applied
to superstring models. 
They provide several new sources of the $\mu$ term, naturally of order the
weak scale,
but also may lead to non--universal soft terms.

\end{abstract}

\thispagestyle{empty}

\leftline{}
\leftline{}
\leftline{FTUAM 97/14}
\leftline{KAIST--TH 97/18}
\leftline{October 1997}

\vskip-15.cm
\rightline{ FTUAM 97/14}
\rightline{KAIST--TH 97/18}
\vskip3in

\end{titlepage}
\newpage
\setcounter{page}{1}

\section{Introduction and Summary}
Superstring theory is by now the only theory which can unify all the known
interactions including gravity. Although this is a remarkable theoretical 
success, experimental evidence is still lacking. 
Since,
unlike philosophy, physics must be subject to experimental test,
if we do not want to regard superstring theory 
in the future 
as a philosophical theory
some experimental evidence must be obtained.
Unfortunately, whereas the  
natural scale of superstring models is 
${\cal O}(M_{Planck})$, 
the `natural' scale of particle accelerators
is 
${\cal O}(1 TeV)$, and therefore 
to obtain any experimental proof seems very difficult.
Nevertheless, if Nature is supersymmetric (SUSY) at the weak scale, 
as many particle physicists
believe, eventually the spectrum of SUSY 
particles will be measured providing us with a possible connection 
with the superhigh--energy world of 
superstring theory.
The main point is that the SUSY spectrum is determined by
the soft SUSY--breaking parameters and these in its turn
can be computed in the context of superstring models.
To compare then the superstring predictions about soft terms with
the experimentally--observed SUSY spectrum would allow us 
to test the goodness of this theory.
This is the path, followed in the last years by several groups, 
that I will try to review briefly in sect. 2 paying special
attention to the constraints deriving from 
flavor changing neutral current (FCNC) phenomena.
In sect. 3 other phenomenological constraints on soft parameters are studied.
In particular, those deriving from charge and color breaking (CCB) 
minima. 
Finally, in sect. 4 I will discuss the effects of pure supergravity (SUGRA) 
loop corrections
on soft parameters and the $\mu$ term and apply them to superstring models.
New sources of the $\mu$ term and FCNC will appear. 

\section{ Soft Parameters from Superstrings and FCNC}
As is well known the soft parameters
can be computed in generic hidden
sector SUGRA models \cite{BIM3}.
They depend on the three functions, $K$, $W$ and $f$ 
which determine the full N=1
SUGRA Lagrangian.
Expanding in powers of the
observable fields $C^{\alpha}$ these are given by
\begin{eqnarray}
K &&=
{\hat K}
+ {\tilde K}_{{\alpha}{\overline{\beta}}}
C^{\alpha} { \bar C^{\overline{\beta}} }\ 
+\frac{1}{4}
{\tilde K}_{{\alpha}{\overline{\beta}}\gamma{\overline {\delta}}}
C^{\alpha} \bar C^{\overline{\beta}}
C^{\gamma} \bar C^{\overline{\delta}}
\nonumber\\ 
&&+
\left[ \frac{1}{2} Z_{{\alpha }{ \beta }}
C^{\alpha}C^{\beta } 
+\frac{1}{2}
{Z}_{{\alpha}\beta{\overline{\gamma}}}
C^{\alpha} C^{\beta} \bar C^{\overline{\gamma}} \right.
\nonumber \\
&& 
+ \left. \frac{1}{6} Z_{{\alpha }{ \beta }\gamma\overline{\delta}}
{C}^{\alpha}
C^{\beta } C^{\gamma} \bar C^{\overline{\delta}} 
+\ h.c.   \right]
+... 
\nonumber 
\\
W &&= {\hat W} +\frac{1}{2}\tilde{\mu}_{{\alpha}{\beta}}{C}^{\alpha}C^{\beta} 
+ \frac{1}{6} \tilde{Y}_{{\alpha}{\beta}{\gamma}}
{C}^{\alpha}C^{\beta}C^{\gamma} 
+...
\nonumber 
\\
f_{ab} &&= 
\hat{f}_{ab}+\frac{1}{2}{\tilde f}_{ab\alpha\beta}C^{\alpha}C^{\beta}
+...
\label{F4}
\end{eqnarray}
where the coefficient functions 
depend upon 
hidden sector fields and
the ellipsis denote the terms 
which are irrelevant for the present calculation.
Although at tree level the cubic and quartic terms in $K$ and the
quadratic terms in $f_{ab}$ are also irrelevant, 
we will see in sect. 4 that when loop effects are included
they may be important for the $\mu$ term and FCNC phenomena.

For example, assuming vanishing cosmological constant, the 
{\it un--normalized}  
scalar masses arising from the
expansion of the (F part of the) 
tree--level SUGRA scalar potential are given in this 
context by 
\begin{eqnarray}
{m}^2_{{\alpha}{\overline{\beta}}} =
m_{3/2}^2{\tilde K_{{\alpha}{\overline{\beta}}}}
- {F}^{m} R_{m{\overline n}\alpha {\overline {\beta}}}
{\overline F}^{\overline n} 
\label{F8}
\end{eqnarray}
with
\begin{eqnarray}
R_{m{\overline n}\alpha {\overline {\beta}}} = 
\partial_m\partial_{\overline{n}}
{\tilde K_{{\alpha}{\overline{\beta}}}}
-\tilde{K}^{\gamma\overline{\delta}}\partial_m\tilde{K}_{\alpha
\overline{\delta}}
\partial_{\overline{n}}\tilde{K}_{\gamma\overline{\beta}}  
\label{ricci}
\end{eqnarray}
where 
$F^m$
denote the hidden field auxiliary components.
Notice that, after normalizing the fields to get canonical kinetic terms,
the first piece in (\ref{F8}) will lead to universal diagonal soft masses
but the second piece will generically induce {\it off--diagonal} contributions.
Actually, universality is a desirable property 
for phenomenological
reasons, particularly to avoid FCNC. 
We will discuss below how string models may get interesting {\it constraints}
from FCNC bounds.

On the other hand, 
from the fermionic part of the tree--level SUGRA Lagrangian,
the {\it un--normalized} Higgsino masses are given by 
\begin{eqnarray}
\mu_{\alpha\beta} =
e^{\hat K/2}
\frac{\hat W^*}{|\hat W|}\tilde {\mu}_{\alpha\beta}+
m_{3/2} Z_{\alpha\beta} - \overline{F}^{\bar n} \partial_{\bar n} 
Z_{\alpha\beta}
\label{mu}
\end{eqnarray}
As is well known the presence of this $\mu$ term,
due to 
the quadratic terms associated with
$Z_{{\alpha}{\beta}}$ and 
$\tilde{\mu}_{{\alpha}{\beta}}$ in (\ref{F4}), 
is 
crucial in order to have correct electroweak symmetry breaking \cite{BIM3}.

The arbitrariness of SUGRA theory, one can think of many possible
SUGRA models (with different $K$, $W$ and $f$) leading to
{\it different} results for the soft terms, can be ameliorated in 
SUGRA
models deriving from superstring theory, where $K$, $f$, and the hidden
sector are more constrained. Let me mention at this point that I will
concentrate here on weakly coupled heterotic string models. 
Although strongly coupled strings (from M-- and F--theory) have recently
been proposed and their phenomenology explored (see \cite{BIM3} for a list
of references), still is too early to
draw conclusions about possible predictions on soft terms. In any case the
results from loop effects by SUGRA interactions that I will discuss
in sect. 4 
can be applied to any superstring theory.
The heterotic string 
models have a natural hidden sector built--in: the dilaton
field $S$ and the moduli fields $T_i$, $U_i$. 
Without specifying the supersymmetry--breaking mechanism, 
just
assuming that the auxiliary
fields of those multiplets are the seed of supersymmetry breaking, 
interesting
predictions for this simple class of models are obtained \cite{BIM3}.

Let us focus first on the very interesting limit where the dilaton $S$
is the source of all the SUSY breaking. At superstring tree level
the dilaton couples
in a universal manner to all particles and therefore, this limit is 
compactification {\it independent}. In particular, since  
the VEVs of the moduli auxiliary fields $F^i$ are vanishing, 
and ${\tilde K_{{\alpha}{\overline{\beta}}}}$ is independent on $S$,
the second piece in (\ref{F8}) is vanishing and therefore,
after normalization, 
the soft scalar masses turn out to be 
universal, 
$m_{\alpha}=m_{3/2}$.
Because of the simplicity of this scenario, the predictions about 
the low--energy spectrum 
are quite precise 
\cite{BIM3}. 

Let us finally remark that although
the soft parameters are universal in this dilaton--dominated SUSY--breaking
limit, this result may be spoiled due to superstring loop effects \cite{louis}. 
In particular,
${\tilde K_{{\alpha}{\overline{\beta}}}}$ receive an $S$--dependent 
contribution and therefore 
the second piece in (\ref{F8})
will be non--vanishing. 
In sect. 4 we will find another source of
non--universality due to generic SUGRA loop effects.

In general the moduli fields, $T_i$, $U_i$, 
may also contribute to SUSY breaking,
i.e. $F^i\neq 0$,
and in that case their effects on soft parameters must also be 
included (see e.g. (\ref{F8})).
Since different compactification schemes give rise to different
expressions for the moduli--dependent part of $K$
(\ref{F4}), 
the computation of the soft parameters will be 
model {\it dependent}. 
To illustrate the main features of mixed dilaton/moduli--dominated 
SUSY breaking, let us 
concentrate first on the simple situation of diagonal moduli and matter 
metrics. This is the case of most
$(0,2)$ symmetric Abelian orbifolds which, 
at superstring tree level, have
${\tilde K}_{\alpha{\overline{\beta}}}=\delta_{\alpha\beta}
\Pi_i(T_i+{\overline T}_i)^{n_{\alpha }^i}$
where $n_{\alpha }^i$ are the 
modular weights of the matter fields $C^{\alpha }$. 
Plugging this result in (\ref{F8})
we get the following
scalar masses after normalizing the observable fields: 
\begin{eqnarray}
{m}^{2}_{\alpha} &&= 
m_{3/2}^2+\sum_i \frac{n_{\alpha}^i}{(T_i+ {\overline{T}_i})^2}
|F^i|^2 
\label{supermasss}
\end{eqnarray}
With this information one can analyze the structure of soft parameters
and in particular the low--energy spectrum \cite{BIM3}.
If we assume that also continuous Wilson--line moduli contribute to
SUSY breaking, the analysis turn out to be more involved since
off--diagonal moduli metrics arise due to the mixing between 
$T_i$, $U_i$ and Wilson lines. 
However, the final formulas for scalar masses are similar to 
(\ref{supermasss}) 
(with some extra contributions due to the Wilson--line auxiliary
fields), and therefore the low--energy spectrum is also similar \cite{kim}.

Notice that the scalar masses (\ref{supermasss}) 
show in general a {\it lack of universality} due to the
modular weight dependence \cite{lust}. 
So, even with diagonal matter metrics, FCNC effects may appear. 
However,
we recall that the low--energy running of the scalar masses has to be
taken into account. In particular, in the squark case, for gluino masses 
heavier than (or of the same order as) the scalar masses at the boundary
scale, there are large flavour--independent gluino loop contributions 
which are the dominant source of scalar masses.
This situation is very common in orbifold models.
The above effect can therefore help in fulfilling the FCNC 
constraints \cite{BIM}.

%
%

Although diagonal metrics is the generic case in most orbifolds, 
there are a few cases, 
$Z_3$, $Z_4$ and
$Z'_6$, where off--diagonal metrics are present in the untwisted sector. 
In particular, at superstring tree level, the 
moduli--dependent part of $K$ 
(\ref{F4}) is given
by: 
$K = -\ln \det(T_{i\bar j}+{\overline T}_{i\bar j}-C^i \bar{C}^{\bar j})$.
This clearly implies that  
the scalar mass eigenvalues will be in general non-degenerate, 
which in turn may induce FCNC \cite{BIMS}. 
This can only be avoided 
under special conditions (for instance, when 
$\hat W$ does not depend on the moduli, a no-scale scenario arises and
the mass eigenvalues vanish). 
The same potential problem is present 
in Calabi--Yau compactifications where off--diagonal metrics is the generic
situation:
$K={\hat K}^T + 
{\hat K}^U +
{\tilde K}^T_{i\overline j} {\bf 27}^i {\bf 27}^{*\overline j} +
{\tilde K}^U_{k\overline l} {\overline {\bf 27}}^k 
{\overline {\bf 27}}^{*\overline l}$
where ${\tilde K}^T_{i\overline j}= 
({\partial}^2 {\hat K}^T/\partial T_i \partial {\overline T}_j)
e^{({\hat K}^U-{\hat K}^T)/3}$, 
${\tilde K}^U_{k\overline l}=
({\partial}^2 {\hat K}^U/\partial U_k \partial {\overline U}_l)
e^{({\hat K}^T - {\hat K}^U)/3}$, 
neglecting $\sigma$ model and instanton corrections
${\hat K}^T=-\ln k_{ijk} 
(T_i+{\overline T}_i)
(T_j+{\overline T}_j)
(T_k+{\overline T}_k)$,
and finally ${\hat K}^U$ is a complicated function of the 
complex structure moduli $U$.
Obviously, using (\ref{F8}), the mass eigenvalues are 
typically non--degenerate \cite{kim2}. 
Only in the special limit when SUSY is broken
by $U$ ($T$)--moduli universality for ${\bf 27}$ ($\overline{{\bf 27}}$) 
representations
is achieved.

\section{ CCB Constraints on Soft Parameters }
We already noticed above 
that the high--energy constraints on the soft 
parameter space obtained from studying superstring models
can be combined with low--energy phenomenological constraints as e.g.
FCNC phenomena.
Obviously, others
constraints arise from demanding that
all the not yet observed particles have masses compatible with 
the experimental bounds. But, in fact, we can go further and impose
the (theoretical) constraint of demanding the no existence of
low--energy
CCB minima deeper than the standard vacuum \cite{mua}.
In the particular case of the
dilaton--dominated scenario, the restrictions coming from the CCB minima
are very strong and the whole parameter space ($m_{3/2}$, $B$, $\mu$)
turns out to be excluded on these grounds \cite{mua2}.
Given these dramatic conclusions, a way--out must be searched.
The simplest possibility is
to assume that also the moduli
fields $T_i$ contribute to SUSY breaking, which is in fact a more general
situation as discussed above. Since then
the soft terms are modified and new free parameters beyond 
$m_{3/2}$ and $B$ appear, possibly some regions in the
parameter space will be allowed. 
Although now the situation is clearly more model dependent, 
a good and
simple starting point might be 
to study the case of orbifolds with diagonal
moduli and matter metrics. Assuming 
that SUSY breaking is equally shared among $T_{i}$'s, i.e. 
the `overall modulus' $T$ scenario, 
basically only one more
free parameter must be added \cite{prepa}. 

\section{ Supergravity Radiative Effects on Soft Parameters and the $\mu$ Term }
The soft parameters (see e.g. (\ref{F8})) and $\mu$ (\ref{mu}) are computed
at the tree level of SUGRA interactions. However, as has been 
shown explicitly in
\cite{choi}, there can be a significant modification in this procedure
due to quadratically divergent SUGRA one--loop effects.
For example, in the case of the $\mu$ term
these quantum corrections 
do not only modify the already known contributions 
from $\tilde{\mu}_{\alpha\beta}$ and $Z_{\alpha\beta}$
in the tree--level matching condition (\ref{mu})
but also provide {\it new} sources of the $\mu$ term, naturally of order
the weak scale. These sources 
depend upon the coefficients
$Z_{\alpha\beta\overline{\gamma}}$ and 
$Z_{\alpha\beta\gamma\overline{\delta}}$ in $K$ 
and also the coefficients $\tilde{f}_{ab\alpha\beta}$
in $f_{ab}$
(\ref{F4}).

Another
interesting feature
of the SUGRA corrections 
is the {\it lack
of universality}. 
If the Riemann tensor (\ref{ricci}) can be factorized as
$R_{m {\overline n}{\alpha}{\overline{\beta}}}=c_{m\overline{n}}\tilde{K}_{\alpha
\overline{\beta}}$, as it happens e.g. in 
no--scale SUGRA models or in the dilaton--dominated scenario explained in
sect. 2,
the tree--level matching conditions would give a universal
soft scalar mass for the {\it normalized} observable fields.
But now it is possible to see that due to the 
SUGRA radiative corrections, particularly due to the contributions
of the pieces
depending upon 
${Z_{{\alpha}\beta{\overline{\gamma}}}}$ and 
${\tilde K_{{\alpha}{\overline{\beta}}\gamma{\overline{\delta}}}}$
in $K$ (\ref{F4}),
the scalar masses will have a generic matrix structure with 
non--degenerate eigenvalues \cite{choi}. 

Let us apply these general results
to  the case 
of superstring effective SUGRA models reviewed in sect. 2.
We will concentrate here on 
(0,2) symmetric Abelian orbifolds with diagonal moduli and matter metrics.
(However our main conclusions will not depend on the particular compactification
scheme used.). 
Although not computed explicitly yet, one can imagine
the following 
modular--invariant form of 
$\tilde{K}_{\alpha\overline{\gamma}\beta\overline{\delta}}:$
${\tilde K}_{\alpha{\overline{\gamma}}\beta{\overline{\delta}}}
=\delta_{\alpha\gamma}\delta_{\beta\delta}  
X_{\alpha\beta}\Pi_i
(T_i+{\overline T}_i)^{n_{\alpha}^i}
\Pi_j
(T_j+{\overline T}_j)^{n_{\beta}^j}$,
where
$X_{\alpha\beta}$
are  constant coefficients of order one.
Using now the SUGRA corrections computed in \cite{choi}, 
the scalar masses (\ref{supermasss}) get an extra contribution
proportional to 
$m_{3/2}^2\sum_{\gamma} 
{X}_{\alpha\gamma}$.
%
One can now see more explicitly the interesting feature 
of the one--loop SUGRA corrections which were discussed briefly above:
even when
the tree--level matching condition 
(\ref{supermasss})
leads to a universal soft mass, which would be 
the case if all $C^{\alpha}$ have the same modular weight
or if all  $F^i=0$ (this corresponds to the case of dilaton-dominated
SUSY breaking),
the SUGRA corrections  
depending upon 
$X_{\alpha\gamma}$
are {\it no longer universal}.

%

\end{document}